\documentclass[preprint2]{aastex63}
\usepackage{amssymb }

\begin{document}

\title{The orbit and density of the Jupiter Trojan satellite system
Eurybates-Queta}

\correspondingauthor{M.E. Brown}
\email{mbrown@caltech.edu}

\author[0000-0002-8255-0545]{Michael E. Brown}
\affiliation{Division of Geological and Planetary Sciences\\
California Institute of Technology\\
Pasadena, CA 91125, USA}
\author[0000-0001-5847-8099]{Harold F$.$ Levison}
\affiliation{Southwest Research Institute \\
 1050 Walnut Street, Suite 300 \\
Boulder, CO 80302, USA}
\author[0000-0002-6013-9384]{Keith S.~Noll}
\affiliation{NASA Goddard Spaceflight Center}
\author[0000-0002-9995-7341]{Richard Binzel}
\affiliation{Massachusetts Institute of Technology}

\author[0000-0003-0854-745X]{Marc W. Buie}
\affiliation{Southwest Research Institute \\
            1050 Walnut Street, Suite 300 \\
            Boulder, CO 80302, USA}
            
\author[0000-0002-8296-6540]{Will Grundy}
\affiliation{Lowell Observatory and Northern Arizona University,
Flagstaff, AZ, USA}
\author[0000-0003-2548-3291]{Simone Marchi}
\affiliation{Southwest Research Institute \\
            1050 Walnut Street, Suite 300 \\
            Boulder, CO 80302, USA}
            \author[0000-0002-5846-716X]{Catherine B.~Olkin}
\affiliation{Southwest Research Institute \\
1050 Walnut Street, Suite 300 \\
Boulder, CO 80302, USA}
\author[0000-0003-4452-8109]{John Spencer}
\affiliation{Southwest Research Institute \\
            1050 Walnut Street, Suite 300 \\
            Boulder, CO 80302, USA}
\author[0000-0003-4909-9542]{Thomas S.~Statler}
\affiliation{NASA Headquarters\\
300 Hidden Figures Way\\
Washington, DC 20546, USA}
\author[0000-0003-0951-7762]{Harold Weaver}
\affiliation{Johns Hopkins University Applied Physics Laboratory}

\begin{abstract}
We report observations of the Jupiter Trojan asteroid (3548) Eurybates
and its satellite Queta with the {\it Hubble Space Telescope} and use these
observations to perform an orbital fit to the system. Queta orbits Eurybates
with a semimajor axis of $2350\pm11$~km at a period of $82.46\pm0.06$ days and an eccentricity of
$0.125\pm0.009$. 
From this orbit we derive a mass of Eurybates of $1.51\pm0.03 \times 10^{17}$ kg, corresponding to an estimated 
density of $1.1\pm0.3$ g cm$^{-3}$, broadly
consistent with densities measured
for other Trojans, C-type asteroids in the outer main 
asteroid belt, and small icy objects from the Kuiper belt. Eurybates is the parent body of the only 
major collisional family among the Jupiter Trojans; its low density suggests
that it is a typical member of the Trojan population. Detailed study of
this system in 2027 with the Lucy spacecraft flyby should allow significant 
insight into collisional processes among what appear to be the icy bodies
of the Trojan belt.
\end{abstract}


\section{Introduction} 
The Jupiter Trojan asteroid (3548) Eurybates is the parent of
the largest known collisional family in the Trojan population
\citep{2011MNRAS.414..565B,2015MNRAS.454.2436V,2015aste.book..297N,2016MNRAS.462.2319R} and has
recently been reported to have a small satellite { with a fractional brightness $\sim$3000 times fainter in }orbit about it
\citep{keith},
now named Queta. In addition, it will be the first Trojan
asteroid to be visited by a spacecraft when the Lucy Discovery
mission arrives on 12 Aug 2027.

Collisional families appear rare in the Trojan population
\citep{2011MNRAS.414..565B,2015MNRAS.454.2436V,2015aste.book..297N,2016MNRAS.462.2319R};
the Eurybates family is the only 
major family,  though a few smaller ones have been suggested. 
Interestingly, Eurybates and its collisional family members
are also among the least red objects in the entire Trojan population
\citep{2007Icar..190..622F, BrownSchemel2021}.
This group is often classified as C-type asteroids in the
\citet{2013Icar..226..723D} taxonomy, for example, 
compared to the P- and D-types found among the rest
of the Jupiter Trojans. For these objects with essentially
featureless spectra out to 2.5~$\mu$m 
and poorly measured albedos, however, it is probably
more informative to simply consider that the Jupiter Trojan
asteroids are bifurcated in color, with a red population (which generally
maps to the
D-type classification) and a less-red population (which generally
maps to the 
P-type classification) \citep{2011AJ....141...25E, 2014AJ....148..112W} 
and that the distribution of colors of Eurybates and its family 
overlap with the less-red colors but skew blueward \citep{2010Icar..209..586D,BrownSchemel2021}.

The small number of collisional families among the Trojans
could  perhaps suggest compositional and
strength differences between the Jupiter Trojan asteroids 
and the main belt asteroid population \citep{2009Natur.460..364L}.
One possibility for the unique family and atypical colors of
Eurybates could be that this object is an interloper, with
a composition more similar to main belt asteroids than to the
potentially ice-rich Jupiter Trojan population. 

Like collisional families, known satellites also appear rare 
among the Jupiter Trojan asteroid population, with Eurybates only the fourth
Jupiter Trojan to have had a satellite directly imaged \citep{keith},
though it is clear that some of this paucity is
an observational bias owing to their greater distance than 
main belt asteroids. It seems likely that, for Eurybates, the
existence of a satellite and of a collisional family are related,
and that both trace back to a catastrophic impact at some earlier
time \citep{2004Icar..167..382D}. The existence of this satellite
allows us to measure the density of Eurybates and potentially help
understand the the critical question of the relationship of
Eurybates to the remainder of the Jupiter Trojans and to the
outer solar system.

\section{Observations and data reduction}
The Eurybates system was observed 13 times between 12 Sep 2018 and 12 Feb 2021 (Table~1). The satellite
was first detected in the initial pair of observations on 12 and 14 Sep 2018 and followup 
to confirm the existence of the satellite and determine the orbit began on 11 Dec 2019.
Details of the observing strategy are given in \citet{keith}.
The first successful re-aquisition occured on 3 Jan 2020, just before Eurybates became
unobservable due to solar avoidance. The satellite was again detected at the 
first opportunity post-solar avoidance on 19 July 2020 and then was not detected 
at the next two attempts. The next detection occurred on 12 October 2020, by which time
the orbit solution had converged sufficiently that the next 4 attempts were all successful.

\begin{deluxetable*}{ccccc}
\tablecaption{Observing circumstances.}

\tablehead{
 \colhead{date\tablenotemark{a}} & \colhead{filter} & \colhead{int. time} & \colhead{R} & \colhead{$\Delta$} \\
 \colhead{(JD)} & \colhead{} &  \colhead{(sec)} & \colhead{(AU)} & \colhead{(AU)}
 }
   
\startdata
2458373.9064  & F555W & 4x350  &    5.371    &  4.599\\
2458375.8933  & F555W & 4x350 &      5.370   &    4.618\\
2458828.7768  & F350LP & 6x350 &         5.074 &      5.122\\
2458839.4619   & F350LP & 6x350 &       5.067 &      5.281\\
2458851.8149 & F350LP & 6x350 &  5.059   & 5.450\\
2459050.3613 & F350LP &  8x220 &  4.934  & 4.754\\
2459065.2545  & F350LP & 8x220  & 4.925 & 4.514\\
2459112.1269  & F350LP & 8x220 & 4.899   & 3.975\\
2459135.2974  & F606W & 2x350 &      4.887 &     3.892\\
             &F814W & 2x480 \\
2459139.1369  &F606W & 2x350&      4.885 &      3.895\\
            &F814W & 2x480 \\
2459177.8612 & F606W   & 2x350&     4.864&       4.145\\
            & F814W & 2x480 \\
2459255.7736 &F350LP & 6x350 &   4.828  &  5.264\\
2459258.1564 &F350LP & 6x350 &   4.827 & 5.292\\
\enddata
\tablenotetext{a}{midpoint of all exposures}
\end{deluxetable*}

\begin{deluxetable*}{clrrrcr}
\tablecaption{Measured position of Queta with respect to Eurybates}

\tablehead{\colhead{observation} &
 \colhead{JD} & \colhead{date} & \colhead{$\Delta$RA} & \colhead{$\Delta$dec} & \colhead{$\sigma$\tablenotemark{a}}& \colhead{$\xi$\tablenotemark{b}} \\
 \colhead{number} & \colhead{(JD)} & & \colhead{(mas)} &  \colhead{(mas)} & \colhead{(mas)} & \colhead{}
 }
   
\startdata
1 &2458373.9064& 12 Sep 2018 & -548   &  103 &   11 & 1.23\\
2 &2458375.8933 &14 Sep 2018 & -486   &  67  &  16 & 0.03 \\
3 &2458828.7768	&11 Dec 2019& - & - & - & -	\\
4 &2458839.4619	&21 Dec 2019&- &-& -& - \\
5 &2458851.8149 &3 Jan 2020&-405  &   408  & 5 & 0.70 \\
6&2459050.3613 &19 July 2020& 432 &	460	&3 & 1.70 \\
7&2459065.2545	&3 Aug 2020& -& -& -& - \\
8&2459112.1269	&19 Sep 2020&-& -& -& - \\
9&2459135.2974	&12 Oct 2020&505&	546&	5 & 1.43 \\
10&2459139.1369	&16 Oct 2020&439 &  487 &   15 & 0.57 \\
11&2459177.8612	&22 Nov 2020&-552 &  -605 &  5   & 1.35 \\
12&2459255.7736	&10 Feb 2021&-443&	-462	&5 & 2.27 \\
13&2459258.1564	&12 Feb 2021& -437&	-476	&4 & 1.50 \\
\enddata
\tablenotetext{a}{The positional uncertainty, assumed to be identical in RA and dec.}
\tablenotetext{b}{The distance from the predicted position divided by the uncertainty}
\end{deluxetable*}

Observations were obtained in a variety of passbands, depending on the expected brightness of
Eurybates at the time of observation. The initial observations were obtained with
F555W. For followup observations we either
used F350LP, when Eurybates was more distant from the Earth and thus fainter,
or a combination of F606W and F814W when Eurybates was brighter and closer.
In all of the followup observations the primary is saturated in order
to get better signal-to-noise on the satellite (additionally unsaturated
short exposures were also obtained, but Queta is not detected in these
so they were not used in
this analysis).

For each set of observations, data reduction proceeds identically. 
All data reduction steps are performed on the flat-fielded images provided
by the HST pipeline (the ``\verb|*_flt.fits|'' files from the archive); no use is made of the re-sampled
geometrically rectified images. First, the 
saturated regions and any regions affected by bleeding of charge from these 
saturated regions along the column are automatically masked;
in practice we conservatively mask everything at approximately 70\% of
the expected saturation level and higher. Next, each image is
manually inspected and pixels with obvious cosmic ray hits are added to the mask.
Low level cosmic rays are always present and likely go unmasked. These events will
contribute to our final uncertainties. 

{ We fit the positions of Eurybates and Queta in two steps: first
we determine the point-spread function (PSF) of the telescope
and the precise position of Eurybates, and then 
we fix the previous two parameters and
fit the brightness of Eurybates and the brightness and position
of Queta to a more localized set of data. 
This two-step process allows the final astrometry of
Queta to be unaffected by the details of
the interior (or exterior) part of the PSF, where the largest discrepancies between the 
model and data often exist.} 
In the first step of our fitting, we create a model 
point-spread function (PSF) using TinyTim \citep{Krist1993} for the
appropriate filter and pixel position. We find that the true PSF of the observations
is more extended than any of the model PSFs, so we convolve the PSF with a symmetric two-dimensional
Gaussian function. We perform a least-squares minimization of all of the unmasked pixels using the IDL implementation
of MPFIT \citep{2009ASPC..411..251M},
allowing the position of Eurybates, the brightness of Eurybates, and the width of the
Gaussian kernel to vary. Even with saturated observations, we find that we can fit the
position of Eurybates quite reproducibly { (as demonstrated by 
the small scatter in final calculated value of the offset position
from Eurybates to Queta)}. We next mask all pixels either within 12 or beyond 25 pixels of Eurybates. 
This step leaves us with an annulus that includes
Queta and the wings of the PSF of Eurybates.
In this second step, we fix the PSF and the 
position of Eurybates to that
previously determined, and perform a least-squares fit allowing the position of
Queta to vary, as well as the flux of Eurybates and Queta. 
We average all of the observations from a visit to 
determine the astrometric offset at the mean time of the
observation.
It is difficult to estimate
uncertainties from such a non-linear process {\it a priori} so we use the dispersion of
the results from multiple measurements during a single HST
visit
as a proxy for uncertainty of the mean. We conservatively assume that the uncertainties
on the measurement in both dimensions of the detector are the same and calculate the
uncertainty as the combined standard deviation of the mean in both dimension. In two
cases (14 Sep 2018, 3 Jan 2020) 
a cosmic ray landed close enough to Queta on
one of the images to affect the measurement, so that individual
observation was discarded from the analysis of that visit. 
All pixel positions
and uncertainties are converted to angles on the sky using the distortion
lookup tables provided in the image headers.
Uncertainties range from 3 mas for observations where the satellite is well-separated and
8 observations were obtained to 16 mas for the observation where Queta was closest
to Eurybates and only 3 non-cosmic ray affected observations were obtained.
The final astrometry is presented in Table 2.
\begin{figure*}
\plotone{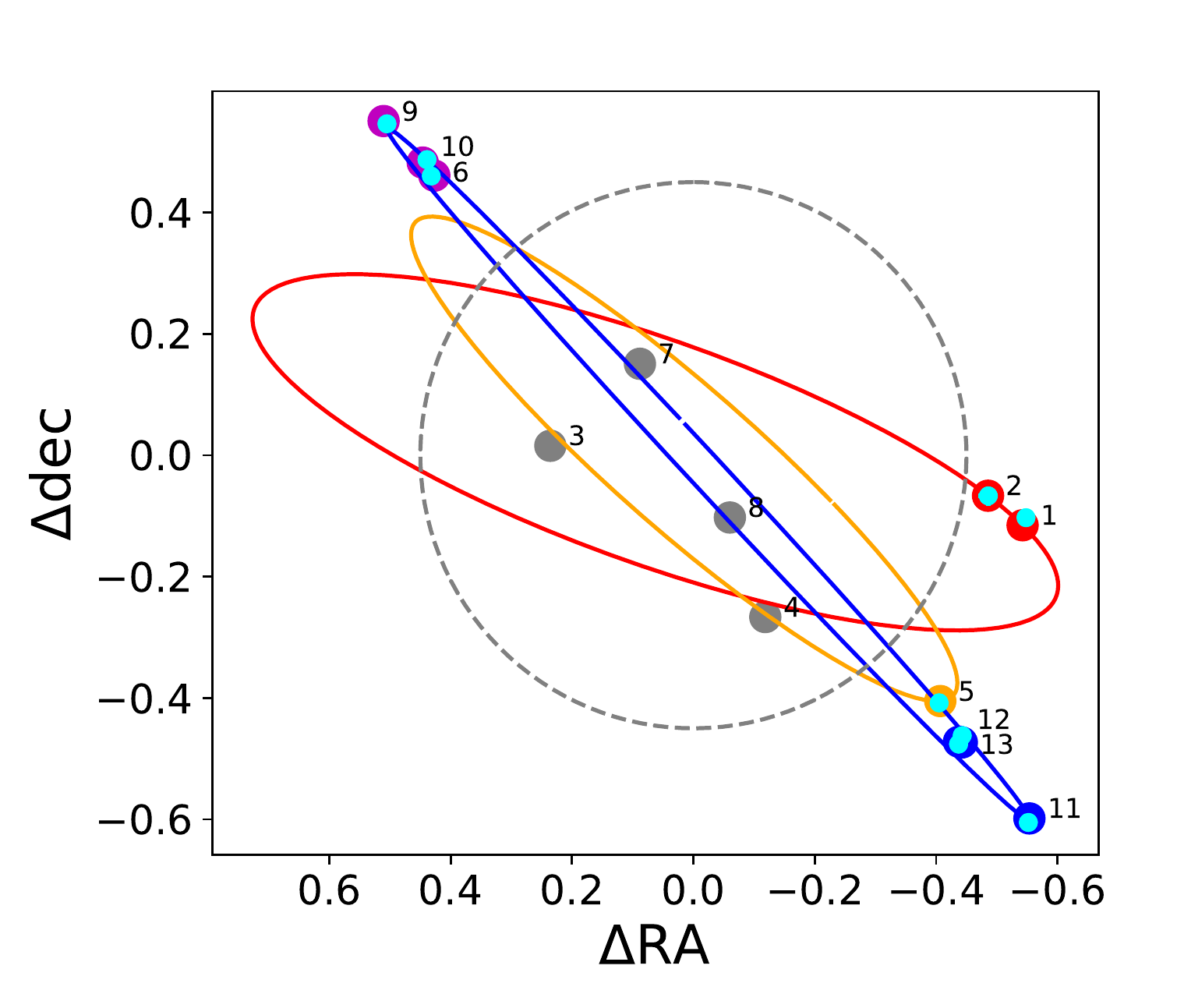}
\caption{{The measured astrometric position of Queta (small
cyan circles) versus the position predicted from the least-squares
fit (large circles),
in units of arcseconds. 
Inside of the
dashed grey circle, Queta is too faint to be detected over the wings of the PSF of Eurybates, and the grey points show the predicted position
of Queta at the times of the non-detections. The numbered labels
refer to the observation number of Table 2.
In most cases the uncertainty in the measured position
is smaller than the size of the symbol. Projections of the orbit at three separate epochs are
shown. The red shows the projection at the epoch of the first two detections (also in red)
The yellow projection shows the orbit at
the moment of the 3 Jan 2020 re-acquisition of the satellite (yellow point). The predicted
position of Queta at the times of the two non-detections
previous to the re-acquisition can be seen in grey along the projection of the yellow orbit at positions 3 and 4.
The blue orbit projection shows the nearly-edge on configuration on 24 Nov 2020 with the
three last detections shown as blue points. The purple predicted points are from Jul and Oct 2020.}
}
\end{figure*}

\section{Orbit fit}
We initially had no knowledge that Earth was 
passing through the orbital plane of the system and thus the viewing geometry was changing rapidly. 
Much effort was thus expended trying to understand the geometry of the orbit  during the period from the first
re-acquisition on 3 January 2020 until a good solution { -- accounting for
all of the detections and also the non-detections -- }was finally obtained 
after the 19 Sep 2020 non-detection, 
Here, however, we simply detail the final orbit fitting. 

The orbit is initially fit with a
least-squares minimization, again using MPFIT,  where all
parameters are allowed to vary. We fit for semimajor axis, 
$a$, orbital period, $P$, inclination (with
respect to the ecliptic), $i$, the eccentricity vector [$e\sin\omega, e\cos \omega$],
the longitude of ascending node, $\Omega$, and the mean anomaly at the epoch of the first observation, $M$. Neither orbital evolution nor external perturbations 
from the Sun are considered, though we revisit this issue later. Figure 1 shows the observed 
astrometric positions of Queta as well as the positions predicted from the least-squares
fit. The least-squares fit has a $\chi^2$ of 8.2 (note that we remove the non-detections from
this fit, but verify that all of the predicted positions at the time of non-detection would
indeed have been undetectable). Given the 9 detections (with both $\Delta$RA and $\Delta$dec)
and the 7 fit parameters, this value corresponds to a reduced $\chi^2$ of 0.75, suggesting
that our  approach to the uncertainties is not unreasonable.

To better understand the range of uncertainties and correlations among the parameters,
we take the initial least-squares fit as the starting position in a Markov Chain Monte Carlo
(MCMC) model. We use the Python package {\it emcee} \citep{2013_Foreman-Mackey} which implements 
the \citet{ISI:000282653600004} affine-invariant MCMC ensemble sampler,
taking uniform priors in the same fit parameters as used above. 
{ While the least-squares fit shows that the non-detections came at
times when the satellite was too close for detection,
to ensure that the MCMC properly penalizes orbits where these
points should have been observed, we include the times of the 
the non-detections in the likelihood calculation
and impose a likelihood penalty if the predicted
position is more than 0.45 arcseconds  from Eurybates
(in practice the predicted positions are all well
inside this radius so they have little effect on the final results).  }
We run a series of 100 separate chains and collect 10,000 samples,
approximately 100 times longer than the autocorrelation timescale of the chain, as determined after-the-fact.
{ For our final results we discard the initial 2000 samples
to ensure that our samples are independent of the initial 
parameters
and take every tenth
sample from the chains to select a statistically independent 
set of samples of the distribution. These samples 
approximate the posterior distribution of the orbital
elements.} 
In these samples, all parameters have approximately Gaussian
marginalized distributions, with some correlation between 
different parameters. As the uncertainties are small, however, 
in Table 3 we
simply give the median and 16th and 84th percentiles as the
$\pm 1\sigma$ range of the parameters.
\begin{deluxetable}{lcc}[htp]
\tablecaption{Parameters of the Eurybates-Queta system\tablenotemark{a}}
\tablehead{}
\startdata
primary parameters \\
$T$ & $82.46\pm 0.06$ & days\\
$a$ & $2350\pm 11$& km\\
$e\sin\omega$&$0.10\pm 0.01$\\
$e\cos\omega$&$0.075\pm 0.004$\\
$i$ &$155.0\pm 0.2$ & deg\\
$\Omega$ &$207\pm 1$ & deg\\
$M$ & $345\pm 3$ & deg\\
epoch & 2458373.9064 \\
\\
derived parameters \\
$e$ & $0.125\pm0.009$\\
$\omega$ & $53\pm 3$ & deg\\
mass& $1.51\pm0.03 \times 10^{17}$& kg \\
density & $1.1\pm 0.3$ & g cm$^{-3}$ \\
$a/R_H\tablenotemark{b}$ & $0.104\pm0.002$
\enddata
\tablenotetext{a}{Referenced to J2000 ecliptic}
\tablenotetext{b}{semimajor axis as fraction of the Eurybates Hill radius}

\end{deluxetable}

From the semimajor axis and period we calculate a mass of Eurybates of $1.51\pm0.03 \times 10^{17}$~kg (the contribution to the system mass of the satellite is  negligible).
The volume of Eurybates has large uncertainties. WISE observations and modeling 
put the effective diameter at $63.9\pm0.3$~km. Correcting
the WISE-assumed absolute magnitude of $H_{\rm V}=9.8$ to
the value of 10.01$\pm$0.02 measured by \citet{2021PSJ.....2...40S}
changes the retrieved diameter by an amount smaller than
the listed uncertainty (though it does drop the best-fit
albedo from 0.052 to 0.043). Systematic errors in the WISE
diameters are difficult to assess, but they include
difficulties in the color corrections for cold targets
such as Jupiter Trojans \citep{2010AJ....140.1868W}, uncertainties owing to the simplistic thermal modeling used \citep{2012ApJ...759...49G}, and the inability
to account for concave surfaces. 
We thus -- perhaps conservatively -- assign 10\% uncertainties
to the WISE diameter. An ongoing occultation campaign \citep{2020DPS....5240102B}
could provide significantly more accurate results
prior to the Lucy flyby.
With our assumed volume of $1.4\pm0.4 \times 10^{14}$ m$^3$, the bulk
density of Eurybates is $1.1\pm0.3$~g~cm$^{-3}$.
The results show that Queta is on a low-eccentricity retrograde
(with respect to the ecliptic) orbit with a moderately long period around a low-density primary.

\section{Orbital evolution}
Queta is widely separated from Eurybates, with a semimajor axis
that is 0.104 times that of the Hill radius of Eurybates, so the satellite 
orbit could be
subject to orbital evolution induced by the Sun.  We investigate
this possibility 
though a long-term dynamical integration in which we use a Wisdom-Holman integrator \citep{1991AJ....102.1528W} to explicitly integrate the full Sun-Eurybates-Queta
system. As an illustration,
Figure~2 shows the evolution of 
Queta's orbit about
Eurybates for the next 3000 years.   { To better understand the
evolution, the orbital parameters are shown with respect to 
the plane of the orbit of Eurybates, rather than in the ecliptic 
coordinate system of Table 3.} The large
oscillations in eccentricity are driven by Kozai { forcing} driven
by the gravitational effects of the Sun \citep{1962AJ.....67..591K}.  In particular,
the solar perturbation causes coupled oscillations between
eccentricity ($e$) and inclination ($i$) such that the magnitude of
the component of angular momentum in the direction perpendicular to
the plane of Eurybates' heliocentric orbit, $l_z = a \sqrt{1-e^2}
\cos{(i)}$, is constant.  In this { orbital integration},
the fact that the argument of pericenter
($\omega$) is circulating indicates that Queta is not in the Kozai
resonance itself.  { Nonetheless}, the observed evolution is induced by 
the resonance islands at
$\omega = 90^\circ$ and $270^\circ$, which force the eccentricity to
become large as $\omega$ sweeps by.

\begin{figure}[htp]
\epsscale{1.1}
\plotone{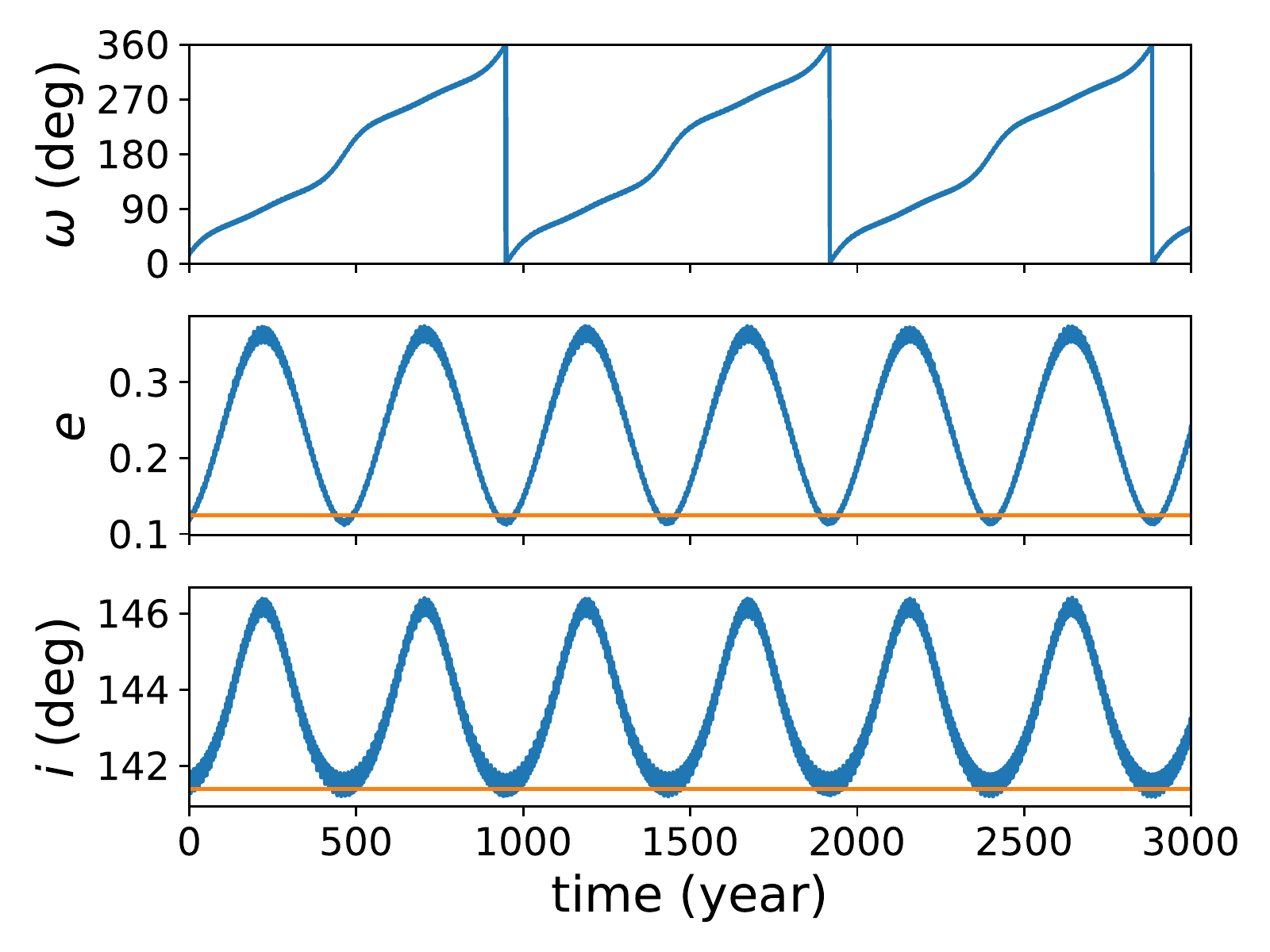}
\caption{\label{fig:orbit}  The temporal evolution
  of Queta's orbit, shown in a reference system defined by
  the plane of the heliocentric orbit of Eurybates, so as
  to illustrate { the Kozai-induced  evolution}. At the top is the
  argument of pericenter, $\omega$, which would librate near 90$^\circ$ or
  $270^\circ$ for an object in the Kozai resonance, { rather than circulate, as seen 
  here}.  The
  middle panel shows eccentricity, $e$, with
  Queta's current observed value of $0.125$ shown by the horizontal
  line.  The inclination, $i$, with respect to the heliocentric
  orbit of Eurybates, is shown on the bottom,{  with the current value of
  139.4$^\circ$ shown.}}
\end{figure}
\begin{figure*}[htp]
\epsscale{.8}
\plotone{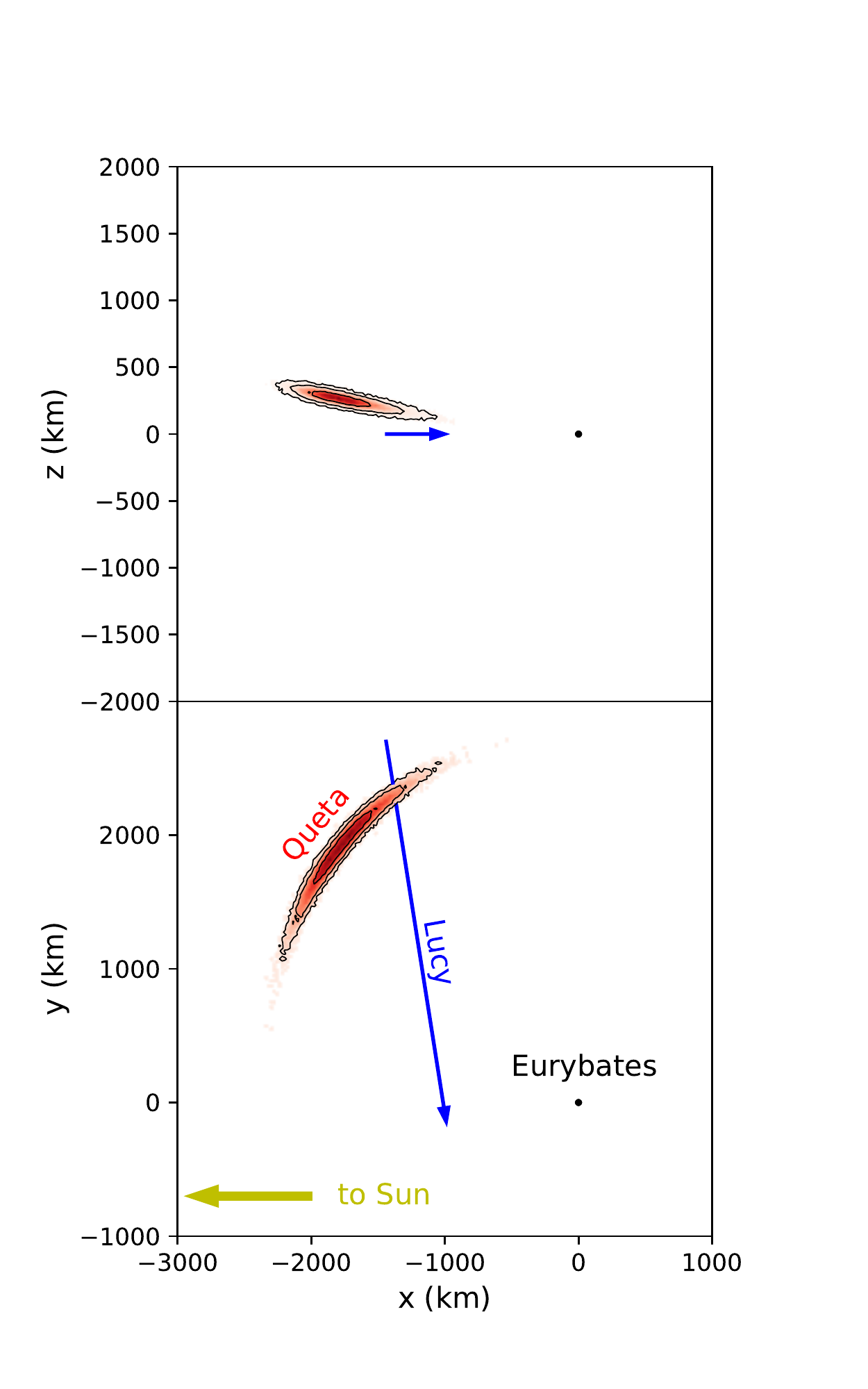}
  \caption{\label{fig:enc}  The probability distribution function
  of the
    location of Queta with respect to Eurybates at time of the Lucy encounter.
    The contours shown are the 50\%, 90\%, and 99\% confidence intervals.
    These data are shown in a
    coordinate system centered on Eurybates, with the Sun on the $-X$
    axis.  Lucy will be targeted to cross the Eurybates-Sun line, 
    thus we define a coordinate system where Lucy always remains
    on the $X$-$Y$ plane. The blue vector shows the path
    of the spacecraft during  500 seconds of the encounter.
     }
\end{figure*}
Assuming no other perturbations, the current orbit should oscillate in eccentricity
from 0.11 to 0.37 and in inclination from 142 to 146$^\circ$
(in the plane of the heliocentric orbit of Eurybates) with a timescale of about 500 years.
The expected orbital evolution over the 29 month span of our observations
is smaller than the uncertainties on our orbital elements, so our orbital
fits, which do not include the effect of the Sun, remain valid. With continued
monitoring of the orbit, however, the orbital elements should get sufficiently
precise that a full fit including the gravitational effects of the Sun will
have to be performed. 

Additional perturbations to the orbit could occur if there are
other satellites in the system. Indeed, an additional satellite
   could disrupt the Kozai resonance, thereby stifling the large
   eccentricity oscillations. Both (87) Sylvia and (107) Camilla
are outer main belt asteroids associated with collisional 
families that have at
least two satellites. In the Kuiper belt, Haumea is the parent of
the only known collisional family and also has at least two satellites. While 
an interior satellite would be difficult to directly image
prior to the Lucy encounter, if such a satellite exists,
its dynamical effects could
perhaps eventually be detected by perturbations to the orbit of Queta or it could be detected through an occultation.

\section{Lucy flyby}

NASA's Lucy mission will perform a close flyby of Eurybates on August
12, 2027.  At closest approach, Lucy will fly within $\sim\!1000$ km
of Eurybates, well within the orbit of Queta. In order
to determine the location of Queta at the time of the encounter,
we take 3000 randomly selected samples from our MCMC chain and propagate
the orbits forward until the time of the encounter.
The span of this extrapolation is sufficiently large that we include
the Sun's effects in this integration, though we find that this step
still makes little difference in the final predictions. The probability distribution
function of the predicted position of Queta is shown in
Figure 3.

The closest approach distances between Queta and Lucy range
from 116 to 1262 km.  The average value of the close approach
distance is 513 km, while the mode of the distribution is 460 km.
Unfortunately, the geometry of the encounter is such that there is
only a 3\% chance that Queta's phase angle will be less than
$90^\circ$ at the moment of close approach, meaning that
the Lucy spacecraft will most likely pass over the unilluminated hemisphere.
Queta will, however, be able to be imaged as the spacecraft approaches the
system.

\section{Discussion}
The low density of Eurybates is consistent with the 0.8 g cm$^{-3}$ of
the Patroclus-Menoetius binary estimated from the orbit-derived
system mass and a volume determined from stellar occultations
 \citep{2015AJ....149..113B}.
 The system mass of Hektor is well known from the orbit of its satellite \citep{2014ApJ...783L..37M}, but estimates of 
 bulk density vary from 1.0 to 2.5 g cm$^{-3}$ due to differences in volume estimates for the bilobed primary \citep{2015Icar..245...64D}. Outside of the Trojans, Eurybates’ density falls in the bottom of the range determined for nine C-complex objects in the main asteroid belt where the system mass has been determined from satellite orbits or spacecraft flyby.  Bulk densities determined for these systems range from 0.7 -  1.8 g cm$^{-3}$ with a mean of 1.4 g cm$^{-3}$ \citep{1999Icar..140...17T, 2015aste.book..355M}. The effects
 of porosity on the densities of these objects is unknown, but
 the low density of Eurybates suggests that it and other low-density asteroids could be derived from the icy population of planetesimals that formed beyond the giant planets for which similar low densities are measured \citep{2005Natur.435..462M}.     
 
The low density of Eurybates suggests that it is consistent with being a typical member 
of the  Jupiter Trojan population, rather than a unique
interloper. If the original 
Eurybates parent body
was indeed a typical Jupiter Trojan asteroid then it is plausible
that the impact itself has led to the bluer-than-usual colors for
Eurybates and the family members. In the hypothesis of \citet{2016AJ....152...90W}, catastrophic collisions will 
remove the original irradiation crust of a 
Jupiter Trojans, and the new irradiation crust formed at
5 AU will be significantly less red than the original. 
Family members formed in this manner would suffer the same fate
and have similarly less-red colors. This hypothesis was
originally developed to explain the two color populations in the
Jupiter Trojans and the observations that the smaller,
thus more collisionally active, objects in the Jupiter Trojan
population are predominantly less red \citep{2014AJ....148..112W,2015AJ....150..174W};
its consistency with the properties of Eurybates and its family is
encouraging. The ability to study this system in detail with
the Lucy spacecraft will allow unique insights into 
the catastrophic impacts and their aftermath in the
icy bodies of the Jupiter Trojan population.

\acknowledgements

This research is based on observations made with the NASA/ESA Hubble Space Telescope obtained from the Space Telescope Science Institute, which is operated by AURA under NASA contract NAS 5-26555. These observations are associated with programs GO-15144 and GO-16056.

\software{MPFIT \citep{2009ASPC..411..251M}; emcee \citep{2013_Foreman-Mackey}}

\end{document}